\documentclass[aps,prb,superscriptaddress,showpacs,twocolumn]{revtex4}

\usepackage{amsmath,amsfonts,amssymb,graphics,graphicx,epsfig,color,times,indentfirst,layout}

\begin{document}

\title{Quantum nondemolition measurements of a flux qubit coupled to a noisy detector}

\author{Wei Jiang}

\affiliation{National Laboratory of Solid State Microstructures and Department of Physics, Nanjing University, Nanjing 210093, China}

\author{Yang Yu}
\email{ yuyang@nju.edu.cn}
\affiliation{National Laboratory of Solid State Microstructures and Department of Physics, Nanjing University, Nanjing 210093, China}

\affiliation{Lab. of Quantum Opt-electronics, Southwest Jiaotong University, Chengdu 610031, China}

\author{Lianfu Wei}
\email{lfwei@swjtu.edu.cn}
\affiliation{Lab. of Quantum Opt-electronics, Southwest Jiaotong University, Chengdu 610031, China}

\begin{abstract}
We theoretically study the measurement-induced dephasing caused by back action noise in quantum nondemolition measurements of a superconducting flux qubit which is coupled to a superconducting quantum interference device (SQUID). Our analytical results indicate that information on qubit flows from qubit to detector, while quantum fluctuations which may cause dephasing of the qubit also inject to qubit. Furthermore, the measurement probability is frequency dependent in a short time scale and has a close relationship with the measurement-induced dephasing. When the detuning between driven and bare resonator equals coupling strength, we will access the state of qubit more easily. In other words, we obtain the maximum measurement rate. Finally, we analyzed mixed effect caused by coupling between non-diagonal term and external variable. We found that the initial information of qubit is destroyed due to quantum tunneling between the qubit states.
\end{abstract}


\maketitle
\section{INTRODUCTION}

The concept of quantum nondemolition (QND) measurement which can be used for detection of weak force acting on system such as gravitational wave had been introduced to allow repeatable detection.\cite{D.F.Walls} The basic requirement of a QND measurement is that a series of measurements of the variable of system should give predictable results.\cite{C.M.Caves} It means that the observable $\hat{A}$ must commute with Hamiltonian $\hat{H}$ that describes interacting system and detector, i.e., $[\hat{A},\hat{H}]=0$. A large number of works associated with QND measurements have been done in quantum optic,\cite{G.J.Milburn,N.Imoto,H.A.Bachor} cavity QED\cite{J.Larson,M. Brune} and circuit QED.\cite{M. Boissonneault}

\begin{figure}
\centering
\includegraphics[width=3.3333in]{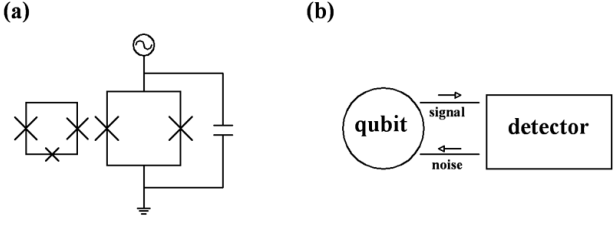}
\caption{(a) Schematic of flux qubit coupled to a SQUID which used for qubit readout as a detector. (b) Schematic of the process of information detraction from qubit and noise injection from one-port detector.
}
\label{fig1}
\end{figure}

Recently, QND measurements have experimentally been achieved on superconducting flux qubit system by investigating the correlation between the results of two consecutive measurements.\cite{A.Lupascu} However,  in general , the measuring apparatus is a mesoscopic system (e.g., dc-SQUID) which is coupled to environment. Therefore, noise can affect discrimination of signal which reflects the information of qubit via coupling between meter and qubit (see Fig.~\ref{fig1}(b)). There exist a larger number of works in mesoscopic system where one is forced to think about the quantum mechanics of detection process,\cite{M.F.Bocko,Y.Makhlin} and about the fundamental quantum limits which constrain the performance of the detector.\cite{A.A.Clerk,D.V.Averin} In practice, noise plays an important role in quantum measurement: quantum noise from the meter acts back on the system, such as qubit, at the same time, the information about the variable conjugate to the measured variable is destroyed. This phenomenon is omnipresent, because information about system is carried away into the surrounding due to indirect environmental coupling via coupling to detector.\cite{L.Viola} For a weak measurement, the detection may be quantum limited that the signal-to-noise of measurement, defined as the ratio of the amplitude of the oscillation line in the output spectrum to background noise, is no more than 4.\cite{D.V.Averin}

In this article, we study the correlation between measurement probability and measurement-induced dephasing. Our model which is used to analyze the interaction between qubit and detector is based on the theory of QND measurement of flux qubit dynamics in a short time.\cite{L.Chirolli} The conclusion shows that the easier we gain information about one variable, the faster we lose information about its conjugate variable. It means that the dc-SQUID can be used as a quantum-limited mesoscopic detector.\cite{A. A. Clerk2} We also study the deviation of QND measurement due to the coupling between non-diagonal term of qubit and variable of detector.  This article is organized as follows. In Sec. II the basic models were introduced firstly, from which we use master equation to discuss the dephasing caused by quantum noise and its correlation with measurement probability distribution. In Sec. III, we discuss the effect of non-diagnose term on QND measurement. We show that the mix effect caused by coupling between non-diagnose term and detector can demolish the information of qubit.

\section{PROBABILITY DISTRIBUTION AND MEASUREMENT-INDUCED DEPHASING}

 The Hamiltonian of system is given by\cite{L.Chirolli}
  \begin{equation}
\hat{H}(t)=\hat{H}_{s}+\hat{H}_{meter}+\hat{H}_{int}+\hat{H}_{drive}(t),
\label{1}
\end{equation}
where $\hat{H_s}$ is the Hamiltonian of qubit which can be considered as a two level system
  \begin{equation}
\hat{H}_{s}=\frac{\epsilon}{2}\hat{\sigma}_{z}+\frac{\Delta}{2}\hat{\sigma}_{x},
\end{equation}
 where $\hat{\sigma}_{z}$ and $\hat{\sigma}_{x}$ are Pauli matrices,  $\epsilon$ and $\Delta$ are energy difference and tunnel splitting of qubit. The Hamiltonian of detector which acts as an oscillator is
 \begin{equation}
 \hat{H}_{meter}=\hbar\omega_{os}\hat a^{\dagger}\hat a.
 \end{equation}
 The Hamiltonian that describes coupling between qubit and oscillator is
 \begin{equation}
 \hat{H}_{int}=\hbar g\hat\sigma_z\hat a^{\dagger}\hat a,
 \end{equation}
 from which we know that the variable that we want to measure is $\hat\sigma_z$. $g$ is coupling strength. In other words, we can conclude the state of qubit from the states of detector according to the value of $\hat\sigma_z$. We assign $\vert 0 \rangle$ to $\sigma_z=1$ and $\vert 1 \rangle$ to $\sigma_z=-1$. The Hamiltonian of external driving of harmonic oscillator is
 \begin{equation}
 \hat{H}_{drive}(t)=f(t)(\hat a+\hat a^\dagger),
 \end{equation}
where $f(t)=2f\cos(\omega_dt)$ is  the driving force with driving frequency $\omega_d$. For an ideal QND measurement, it is necessary that the condition $[\hat{\sigma}_z, \hat{H}]=0$ must be satisfied. However, for superconducting flux qubits, due to the off-diagonal term ${\Delta}\neq0$, the measurement is not perfect QND. In case that $\Delta\ll\epsilon$ is satisfied, $\hat\sigma_z$ still would be treated as a conserved quantity on the time scale determined by $1/\epsilon$ and one expects small deviations from an ideal QND case\cite{L.Chirolli}. After neglecting the off-diagonal term and moving in the rotating frame with frequency $\omega_d$, we obtain the new Hamiltonian
\begin{equation}
\hat{H}=\frac{\epsilon}{2}\hat\sigma_z+\hbar(\delta\omega+g\hat{\sigma}_z)\hat a^\dagger\hat a+f(\hat a+\hat a^\dagger)
\end{equation}
with $\delta\omega=\omega_{os}-\omega_d$. Then we can get a QND measurement from the form of the Hamiltonian.
In order to find the relationship between measurement-induced dephasing and probability distribution, we analyze the impact of noise on QND measurement in part A and make a more quantitative analysis in part B.

\subsection{THE IMPACT OF NOISE ON QND MEASUREMENT}

Based on the model mentioned by Ref. [16], we substitute a Gaussian white noise for the zero point fluctuation because of the fact that noise contributes a large part in the actual measurement.
The distribution of qubit signal $I$ conditioned on the qubit state $\sigma_z=\pm1$ is
\begin{equation}
p(I\vert\hat\sigma_z)=\frac{1}{\sqrt{2\pi S_{II}t}}\mathrm{exp}[\frac{-(x-x_i)}{2S_{II}t}],
\end{equation}
where $x_i(t)=\sqrt{2}\mathrm{Re}[\alpha_i(t)]$, $i=\pm$, $S_{II}$ is noise spectra density. After turning on the interaction with time t, the coherent state qubit-dependent amplitude is found to be\cite{L.Chirolli}
\begin{equation}
\alpha_i(t)=A_ie^{i\phi_i}[1-e^{-i\delta\omega_it-\kappa t/2}]
\end{equation}
with qubit-dependent detuning $\delta\omega_i=\delta\omega-g\sigma_z$ and the qubit-dependent amplitudes and phase given by
\begin{equation}
A_i=\frac{f}{\sqrt{(\delta\omega_i)^2+\kappa^2/4}}
\end{equation}
\begin{equation}
\phi_i=\arctan{(\frac{\delta\omega_i}{\kappa/2})}-\frac{\pi}{2}.
\label{10}
\end{equation}

\begin{figure}
\centering
\includegraphics[width=3.3375in]{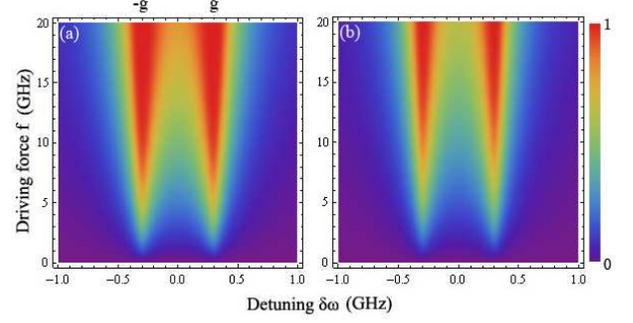}
\caption{ (color online) (a) Probability distribution considering zero point fluctuation for getting the measurement result for 0 and the initial state $\vert0\rangle\langle0\vert$, plotted as a function of detuning $\delta\omega$. (b) Probability distribution considering the back action noise instead of zero point fluctuation with the same parameter. The damping rate of resonator $\kappa=0.1 \mathrm{GHz}$, the couple strength between qubit and oscillator $g=0.3 \mathrm{GHz}$ and the back action noise spectra density $S_{II}=2/\kappa$. }
\label{fig2}
\end{figure}

If we choose a time $t\sim1/\epsilon$, then the signal difference becomes\cite{L.Chirolli}
\begin{equation}
\delta\alpha(t)\approx\sqrt{2}tA,
\end{equation}
where $A=f(e^{2i\phi_0}-e^{2i\phi_1})/\sqrt{2}$. We obtain the outcome probability distribution
\begin{equation}
P(I,t)=\frac{1}{2}[1+I\,\mathrm{erf}(\frac{\delta x(t)}{\sqrt{2S_{II}t}})\langle\hat\sigma_z\rangle_0].
\end{equation}
If we measure a rotated quadrature $\hat X_{\phi}=(ae^{-i\phi}+a^\dagger e^{i\phi})/\sqrt{2}$ and choose the phase $\phi=\arg\mathrm{A}$. The qubit difference becomes $\delta x(t)=\vert A\vert t$, and the probability for the measurement is
\begin{equation}
P(I,t)=\frac{1}{2}[1+I\langle\hat\sigma_z\rangle_0\mathrm{erf}(\frac{\vert A\vert t}{\sqrt{2S_{II}t}})].
\end{equation}

In Fig.~\ref{fig2} we plot the probability of measuring the 0 state under the condition that we consider the contribution of zero point fluctuation (Fig.~\ref{fig2} (a)). We also plot the probability of measuring the 0 state under the condition that we consider the contribution of back action noise instead of the zero point fluctuation during the measurement process (Fig.~\ref{fig2} (b)). Due to the influence of back action noise, we can find that the strong measurement region, compared with the one that only considered the zero point fluctuation, is reduced. It means that we need more time to reach the same measurement strength as the situation that we only consider contribution of the zero point fluctuation.

\subsection{CORRELATION BETWEEN MEASUREMENT PROBABILITY AND DEPHASING RATE}

We choose the initial state of system to be $\rho(0)=\rho_0\otimes|0\rangle\langle0|$, i.e., at time $t=0$ the harmonic oscillator in the vacuum state. Then, we turn on qubit and oscillator interaction, after a period of time, qubit and oscillator entangle with each other. At the same time, we perform a strong measurement of the flux quadrature $\hat{x}=\hat{a}+\hat{a}^\dagger$. As a result, the pointer state (the oscillator state) is projected to the state $|x\rangle\langle x|$. If the measurement is strong enough that the states of oscillator $|x_{\pm}\rangle\langle x_{\pm}|$ which correspond to the qubit states $\vert 0\rangle$ and $\vert 1\rangle$ respectively are orthogonal, the coherence of qubit is destroyed by the measurement. To describe this phenomenon, we must find out the equation of motion that can describes the system state evolution. The master equation used in circuit QED system is valid here.\cite{Gambetta} The qubit density matrix obeys:
\begin{equation}
\dot\rho=\mathcal{L}\rho=-\frac{i}{\hbar}[\hat{H}, \rho]+\kappa\mathcal{D}[\hat{a}]\rho+\gamma_1\mathcal{D}[\hat{\sigma}_-]\rho+\frac{\gamma_2}{2}\mathcal{D}[\hat{\sigma}_z ]\rho,
\label{7}
\end{equation}
where $\mathcal{D}[\hat{L}]$ is damping superoperator defined by $\mathcal{D}[\hat{L}]=(2\hat{L}\rho\hat{L^\dagger}-\hat{L^\dagger}\hat{L}\rho-\rho\hat{L^\dagger}\hat{L})/2$. In above expression, $\kappa$ is decay rate of resonator, $\gamma_1$ is qubit relaxation rate and $\gamma_2$ is dephasing rate.
To figure out the effect induced by measurement due to coupling of qubit to the oscillator, we do not consider the coupling between qubit and environment. Therefore, we neglect relaxation due to coupled to environment and set $\gamma_1=0$, (i.e., for an ideal QND measurement no quantum tunneling occurs between qubit states$\vert0\rangle$ and $\vert1\rangle$ during measurement). Eq.~(\ref{7}) can be solved under the condition that the density matrix with an expansion in positive $\mathnormal{P}$ representation. In Ref.~[10] the solution of the master equation was found to be
\begin{equation}
a_{10}(t)=a_{10}(0)\mathrm{exp}[-i(\epsilon-i\gamma_2)t-i2g\int^t_0\alpha_+(t)\alpha^*_-(t)dt']
\label{15}
\end{equation}
and $a_{01}(t)=a^*_{10}(t)$. The amplitude of coherence state $\alpha_{\pm}(t)$ are determined by
\begin{equation}
\alpha_+(t)=\alpha_{+}^s+\mathrm{exp}[-(\kappa/2+ig+i\delta\omega)t](\alpha_+(0)-\alpha^s_+)
\end{equation}
\\
with $\alpha^s_+=-if/(i\kappa/2+ig+\delta\omega)$, and
\begin{equation}
\alpha_-(t)=\alpha_-^s+\mathrm{exp}[-(\kappa/2-ig+i\delta\omega)t](\alpha_-(0)-\alpha^s_-)
\end{equation}
\\
with $\alpha_-^s=-if/(k/2-ig+i\delta\omega)$, where $\alpha^s_{\pm}$ is the steady coherence state value of the oscillator, $\alpha_+$ and $\alpha_-$ are amplitudes of coherent state $\vert\alpha_+\rangle$ and $\vert\alpha_-\rangle$, which depends on qubit states, $a_{01}$ is the coherence term.

\begin{figure*}
\centering
\includegraphics[width=6.4375in]{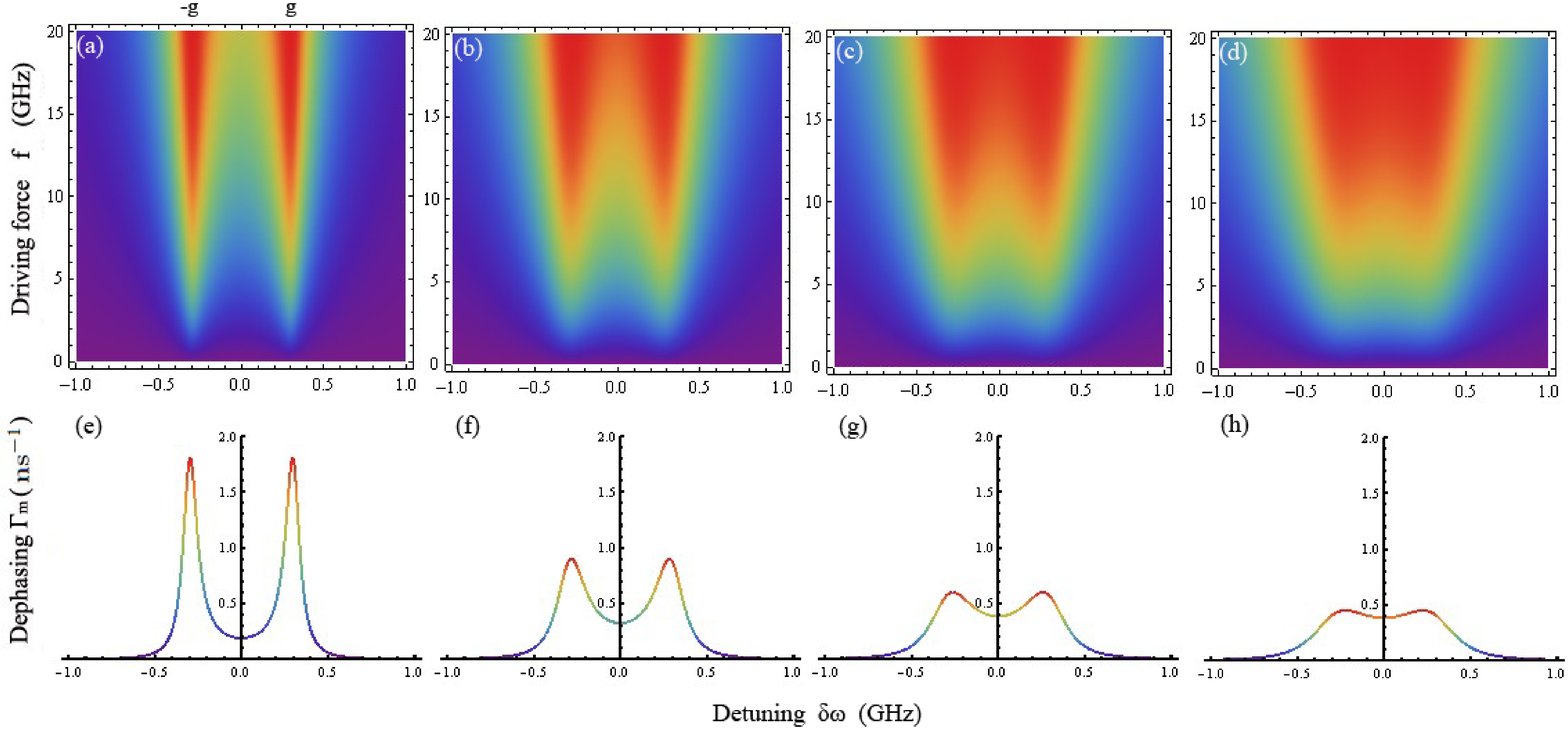}
\caption{(color online)  (a)$\sim$(d) Probability of measuring 0 state for the initial state $\vert0\rangle\langle0\vert$ with damping rate of oscillator $\kappa=0.1~\mathrm{GHz}$, $\kappa=0.2~\mathrm{GHz}$, $\kappa=0.3 \mathrm{GHz}$, $\kappa=0.4 \mathrm{GHz}$, respectively, and qubit coupling strength $g=0.3 \mathrm{GHz}$, $t=0.1 \mathrm{ns}$, noise spectra density $S_{II}=2/\kappa$. (e)$\sim$(h) measurement-induced dephasing rate $\Gamma_m$ as a function of detuning $\delta\omega$ between resonator and drive frequency with driving force $f=1 \mathrm{GHz}$. The parameters of (e)$\sim$(h) are same with those in (a)$\sim$(d), respectively.}
\label{fig3}
\end{figure*}

  We can find that $\gamma_2$ is the intrinsic dephasing rate from Eq.~(\ref{15}) and $\Gamma_m$ is the measurement-induced dephasing rate which is determined by $\alpha_{\pm}^s$ for a project measurement and expressed as following\cite{Gambetta}
\begin{equation}
\Gamma_m=2g\mathrm{Im}(\alpha_+^s\alpha_-^s)=\frac{(n_++n_-)\kappa g^2}{\kappa^2/4+g^2+\delta\omega^2},
\label{17}
\end{equation}
where $n_{\pm}=\vert\alpha_{\pm}^s\vert^2=f^2/[\kappa^2/4+(\delta\omega\pm g)^2]$ is average number of photons in the resonator. We find that for $\delta\omega=\pm\sqrt{g^2+\kappa^2/4}$ we will get a large probability if measure the amplitude of the signal. In other words, when matching one of the frequencies of qubit (i.e., $\delta\omega=\pm g$), the information of qubit is encoded in the amplitude rather than phase which is the conjugate variable with amplitude. At the same time, the measurement-induced dephasing rate will get the maximum value. In Fig.~\ref{fig3}, we plot the probability distribution that measuring 0 state conditioned on initial state is prepared on $\vert0\rangle\langle0\vert$ and measurement-induced dephasing as a function of detuning. When matching one of qubit frequency, both measurement-induced dephasing rate and measurement probability get a maximum value. In the region where $\mathrm{erf}(\vert A\vert t)\approx1$, it corresponds to a strong projective measurement case. At the same time, the two coherent states of oscillator are well separated in phase space due to decay of coherence term at a measurement-induce dephaing rate $\Gamma_m$. In general case, at time $t=0$, the system is in a product state
\begin{equation}
\vert\psi(t=0)\rangle=\frac{1}{2}(\vert0\rangle+\vert1\rangle)\otimes\vert\alpha\rangle,
\end{equation}
where $\vert\alpha\rangle$ is the initial state of oscillator. At time t, the state of system can be written as
\begin{equation}
\vert\psi(t)\rangle=\frac{1}{2}(\vert0\rangle\vert\alpha_+\rangle+\vert1\rangle\vert\alpha_-\rangle).
\end{equation}

Then, if we know oscillator state is $\alpha_i$, the qubit state can be determined exactly. From Eq.~(\ref{17}) we find that the measurement-induced dephasing depends on the overlap of both two oscillator state $\alpha_\pm$\cite{Jay Gambetta2}. In case that the oscillator states $\vert\alpha_-\rangle$ and $\vert\alpha_+\rangle$ are orthogonal, the measurement could be considered as a strong projective measurement, i.e., in the region $\mathrm{erf}(\vert A\vert t)=1$. For the maximum dephasing value, the oscillator states decay to a steady state significantly, which explains the feature of Fig. \ref{fig3} that both measurement-induced dephasing rate and measurement probability get a high value for matching the qubit frequency. In this situation, the qubit state is encoded in the amplitude of coherence rather than phase. It means that the easier we gain information about amplitude, the faster we lose information about phase. The measurement-induced dephasing rate can also be expressed as
\begin{equation}
\vert\langle\alpha_-(t)\vert\alpha_+(t)\rangle\vert=e^{-\Gamma_m t}.
\end{equation}

In addition, as shown in Fig. \ref{fig3}, with the value of damping rate $\kappa$ increasing, both structures of probability distribution and measurement-induced dephasing rate become flat with synchronous tendency, which consistent with the expectation. Increasing of driving force $f$ only change the overall scale of (e)$\sim$(h) rather than the structure, which explains the trend that with the value of driving force increasing, the strong measurement region in (a)$\sim$(h) becomes wider and the measurement probability increases significantly, especially for matching the qubit frequency.

\section{DEVIATION OF A PERFECT QND MEASUREMENT DUE TO MEASUREMENT-INDUCED TUNNELING}

Taking into account the impact of non-diagonal term of qubit, we obtain the system Hamiltonian
\begin{equation}
\hat{H}_s=\frac{\epsilon}{2}\hat\sigma_z+\frac{\Delta}{2}\hat\sigma_x.
\end{equation}
Then we can diagonalize the qubit Hamiltonian $\hat{H}_s$. In new eigenvector space, the system Hamiltonian can be written as
\begin{equation}
\hat{H}_s=\frac{E}{2}\hat\sigma_z,
\end{equation}
meanwhile, the interact Hamiltonian can be written as
\begin{equation}
\hat{H}_{int}=\hbar g\hat\sigma_n\hat{a}^\dagger\hat{a},
\label{29}
\end{equation}
where $E=\sqrt{\epsilon^2+\Delta^2}/2$ is energy splitting of qubit and $n$ is a vector that represents the direction of qubit basis before diagonalized relative to the energy basis
\begin{equation}
\hat\sigma_n=\cos\eta\hat{\sigma}_z+\sin\eta\hat\sigma_x,
\end{equation}
with $\eta=\arctan{(\Delta/\epsilon)}$. The eigenstates are denoted by the superposition of $\vert0\rangle$ and $\vert1\rangle$
\begin{gather}
\vert\uparrow\rangle=\cos\frac{\eta}{2}\vert0\rangle+\sin\frac{\eta}{2}\vert1\rangle\notag\\
\vert\downarrow\rangle=-\sin\frac{\eta}{2}\vert0\rangle+\cos\frac{\eta}{2}\vert1\rangle.
\end{gather}

From Eq. (\ref{29}) we find that fluctuation induced by external noise in particle number $\hat{n}=\hat{a}^\dagger\hat{a}$ can causes transition between qubit states. Therefore, qubit acts as a spectrum analyzer. We separate the interaction Hamiltonian into two parts, one is $\hbar g\cos\eta\hat\sigma_z\hat{a}^\dagger\hat{a} $ and the other is $\hbar g\sin\eta\hat\sigma_x\hat{a}^\dagger\hat{a}$. The first part does not affect the repeatability of QND measurement as discussed in previous sections. The second part will cause ``spin flip'' transition due to the coupling between qubit off-diagonal term and external variable. According to the equation of motion, the time evolution of density matrix is obtained in interaction picture by

\begin{equation}
\dot{\rho}^{I}(t)=\frac{1}{i\hbar}[\hat{H}^I_{int}(t),\rho^{I}(t)].
\label{32}
\end{equation}

We choose the initial state as the a product of qubit and detector, i.e., $\rho(0)=\rho_{0}(0)\otimes\rho_{D}(0)$. Then, tracing Eq. (\ref{32}) over the detector degrees and expanding its right hand to the second order, we obtain the time evolution of reduced density matrix $\rho_{0}$. Eq. (\ref{32}) becomes

\begin{equation}
\begin{split}
\dot{\rho}^{I}_{0}(t)={}&-\frac{1}{\hbar^2}\int_{0}^{t}d\tau\{[\hat\sigma^{I}_{n}(t),\hat{\sigma}^{I}_{n}(t-\tau)\rho^I(t-\tau)]\langle\hat{n}^I(\tau)\hat{n}\rangle_{D}\\
&-[\hat\sigma^{I}_{n}(t),\rho^{I}(t-\tau)\hat\sigma^{I}_{n}(t-\tau)]\langle\hat{n}\hat{n}^I(\tau)\rangle_{D}\}.
\end{split}
\label{33}
\end{equation}

After taking matrix elements between eigenstates of $\hat{H_s}$, we finally obtain the equation of motion that can describe the reduced density matrix evolution
\begin{equation}
\dot\rho^{I}_{kk'}(t)=-\int_{0}^{t}d\tau\sum_{l,l'}M_{kk'll'}\rho^I_{ll'}(t),
\label{34}
\end{equation}
in which we introduced the matrix
\begin{multline}
M_{kk'll'}=\\
g^2\{\langle\hat{n}^I(\tau)\hat{n}\rangle[(\hat\sigma^{I}_n(t)\hat\sigma^I_n(t-\tau))_{kl}\delta_{l'k'}-(\hat\sigma^I_n(t-\tau))_{kl}(\hat\sigma^I_n(t))_{l'k'}]\\
+\langle\hat{n}\hat{n}^I(\tau)\rangle[(\hat\sigma^{I}_n(t-\tau)\hat\sigma^I_n(t))_{l'k'}\delta_{kl}-(\hat\sigma^I_n(t))_{kl}(\hat\sigma^I_n(t-\tau))_{l'k'}]\}.
\end{multline}

To calculate the energy relaxation rate and dephasing rate, we compute the transition rate between qubit eigenstates by computing evolution of the diagonal terms of reduced density matrix $\rho_{11}$ and $\rho_{00}$, and the decay of the off-diagonal term $\rho_{01}$ with the initial state being a $\hat\sigma_x$ eigenstate. The transition rates between the two qubit eigenstates can be derived from Eq. (\ref{34})

\begin{gather}
\Gamma_{\downarrow}=g^2\sin^2\eta S_{nn}(\omega_{01})\notag\\
\Gamma_{\uparrow}=g^2\sin^2\eta S_{nn}(-\omega_{01}),
\end{gather}
where $\Gamma_{\uparrow}$ and $\Gamma_{\downarrow}$ are rates at which qubit is excited from ground state to excited state and decay from excited state to ground state, respectively. $\omega_{01}=E/\hbar$, $S_{nn}(\omega)=\int_{0}^{t}d\tau e^{i\omega\tau}\langle\hat{n}(\tau)\hat{n}(0)\rangle$. If the noise source is in the thermal equilibrium at temperature $T$, the transition rate must satisfy the balance condition $\Gamma_\uparrow/\Gamma_\downarrow=\mathrm{exp}(-\beta\hbar\omega_{01})$ with effective temperature $\beta=1/k_BT_{eff}$. It means that the qubit energy can be absorbed or emitted by detector and the ratio between positive and negative noise spectral density depends on the effective temperature. We can also obtain the expression of dephasing rate $\Gamma_{\phi}$
\begin{equation}
\begin{split}
\Gamma_{\phi}&=\frac{1}{2}g^2\sin^2\eta[S_{nn}(\omega_{01})+S_{nn}(-\omega_{01})]\\
&+g^2\cos^2\eta S_{nn}(\omega=0)\\
&=\frac{\Gamma_{\uparrow}+\Gamma_{\downarrow}}{2}+\gamma_{\phi},
\label{37}
\end{split}
\end{equation}
where $\gamma_{\phi}$ is qubit pure dephasing rate. For the case $\Delta=0$, the measurement is a perfect QND measurement which cannot causes transition between qubit states, in which the dephasing rate is
\begin{equation}
\Gamma_\phi=\gamma_\phi=g^2\int_{-\infty}^{\infty}d\tau\langle\hat{n}^I(\tau)\hat{n}\rangle_D.
\label{38}
\end{equation}
To calculate dephasing rate caused by fluctuation of particle number of oscillator during measurement, we should calculate correlator $\langle\delta\hat n(\tau)\delta\hat{n}\rangle$, where $\delta\hat{n}$ is quantum fluctuations around the mean of photons in the resonator. For a driving resonator with frequency $\omega_d$, the following equation is valid\cite{Blais}
\begin{equation}
\hat{a}(t)=e^{-i \omega_d t}[\alpha+\hat{d}(t)],
\end{equation}
where $\alpha$ is a classic part corresponding to the coherence state $\vert\alpha\rangle$ and $\hat d(t)$ is a quantum part which can annihilate the coherence state to the vacuum state. Then the correlator becomes
\begin{equation}
\langle\delta\hat{n}(\tau)\delta\hat n(0)\rangle=\langle\hat d(\tau)\hat d(0)\rangle=\overline{n}e^{i\delta\omega t-i\frac{\kappa}{2}t},
\end{equation}
where $\overline{n}$ is the average photo number in the resonator. Under the condition that $\delta\omega=0$ and weak coupling $g\ll\kappa$, the measurement induced dephasing rate is
\begin{equation}
\Gamma_m=\kappa\overline{n}\theta_0^2,
\end{equation}
where $\theta_0=\arctan{(2g/\kappa)}\approx2g/\kappa$ is phase shift during measurement derived from Eq.~(\ref{10}), which is identical with Ref.~[10] and Ref.~[19] with a coefficient of expression difference that is caused by the difference of the lower limit of integral. It is quite a contrast to Eq.~(\ref{17}) that has double peaks for matching the frequencies of qubit. With the amplitude of off-diagonal term of density matrix increasing, the pure dephasing rate become weak and relaxation will be dominant.

Form Eq.~(\ref{38}), we find that the decay of coherence is caused by intrinsic dephasing and aggravated by measurement-induced dephasing. In addition, the deviation from the initial value of $\langle\sigma_z\rangle$ become bigger as the time goes by due to transition between the qubit states. Therefore, the state of qubit is acquired; meanwhile, the repeatability of measurement is destroyed. It means that it should not be considered as a QND measurement. Based on this effect, it can be used for ``back action cooling'' in many system, such as a cantilever coupled to a optical cavity,\cite{Marquardt} a noisy qubit coupled to a mechanical resonator,\citet{Y. D. Wang} nanomechanical resonator coupled to a driven superconducting resonantor\cite{Rocheleau} and cooling a micromirror by radiation pressure inside a optical cavity.\cite{Gigan}

\section{CONCLUSION}

We have analyzed the QND measurement of a flux qubit coupled to a noisy detector.
The analytical results reveal that both of the measurement probability and measurement-induced dephasing have a similar trend that their values have peaks for matching one of the qubit frequencies and reduce at the resonance point, at which driving frequency equals the bare harmonic oscillator frequency. With the resonator decay rate increasing, both of the curves of dephasing rate and measurement probability  become flat. Due to the coupling between the non-diagonal term of qubit and variable of detector, the noise of detector perturbs the initial state of qubit. As a result, the transition between the qubit states destroys the initial information of qubit. Therefore, there will be energy exchange between qubit and detector according to the value of effective temperature of noise, which can be used for ``back action'' cooling. Conversely, the behavior of qubit which acts as a spectrum analyzer can also be used for analyzing the property of noise.

\section{ACKNOWLEDGEMENT}

This work was partially supported by the
State Key Program for Basic Research of China (2006CB921801) and NSFC (10725415).

\end{document}